\documentclass[amsmath,amssymb,floatfix]{revtex4}
\usepackage{graphicx}
\usepackage{subfigure}
\usepackage{epstopdf}
\usepackage{amsmath} 
\usepackage{subfigure}
\usepackage{overpic}

\begin{document}
\title{Supporting Information: Short and long time drop dynamics on lubricated substrates}
\author{Andreas Carlson$^{1, 3, 4,}$}
\email{carlson@seas.harvard.edu, hastone@princeton.edu}
\author{Pilnam Kim$^{2,4}$} 
\author{Gustav Amberg$^3$}
\author{Howard A. Stone$^{4,*}$}
\affiliation{$^1$ School of Engineering and Applied Sciences, Harvard University, Cambridge, USA.}
\affiliation{$^2$ Department of Bio and Brain Engineering, Korea Advanced Institute of Science and Technology, Daejeon, South Korea.}
\affiliation{$^3$ Department of Mechanics, Linn\'e Flow Center, Royal Institute of Technology, Stockholm, Sweden.}
\affiliation{$^4$ Department of Mechanical and Aerospace Engineering, Princeton University, Princeton, USA.}
\date{\today}

\maketitle

\section{Determining master curves for the inertial and viscous regime} 
\begin{figure}[ht!]
\centering
\includegraphics[trim=0cm 0cm 0cm 0cm, width=1.0\linewidth]{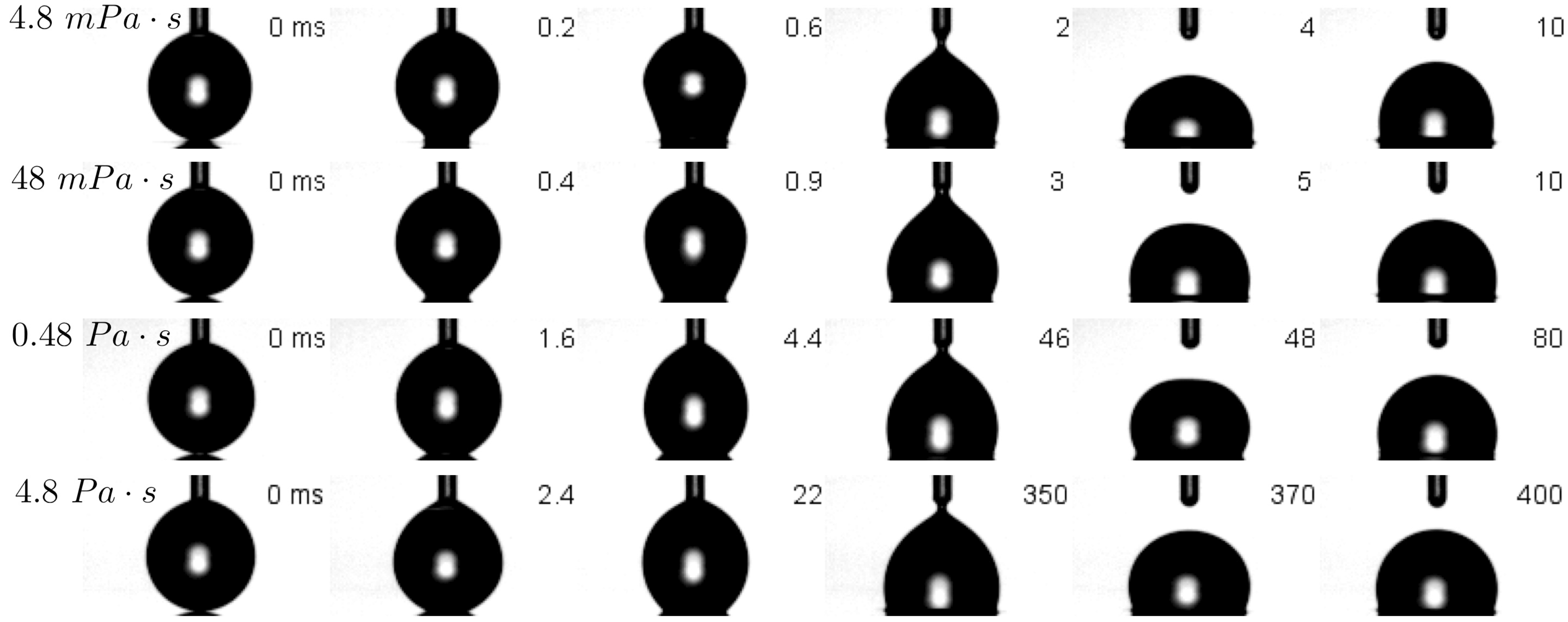}
\caption{Time history of drop shapes as a function of film viscosity ($\mu_o$), indicated to the left. At the point of detachment from the needle the drop shape is independent of the film viscosity.}
\label{fig:Appdet}
\end{figure}

\begin{figure}[!ht]
\centering
\subfigure[ ]{
{\includegraphics[width=0.450\linewidth]{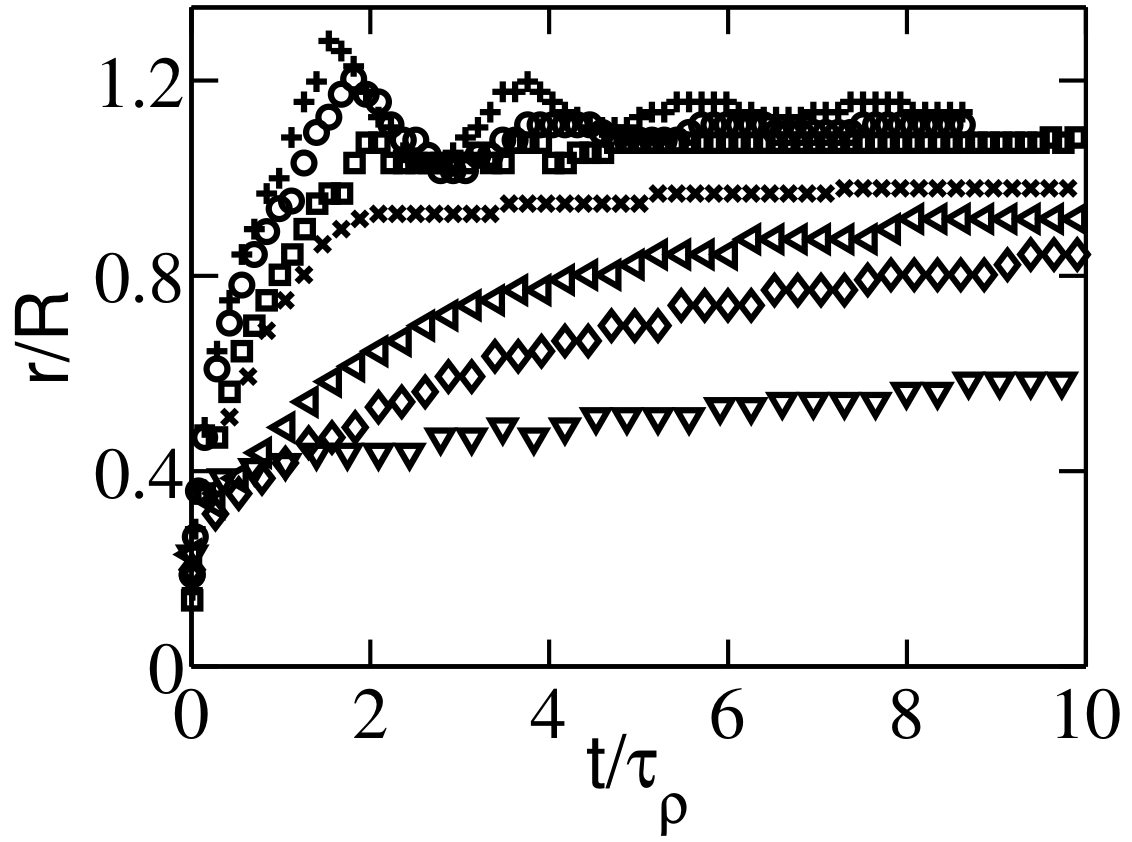}}
\label{fig:2b}
}
\centering
\subfigure[ ]{
{\includegraphics[width=0.450\linewidth]{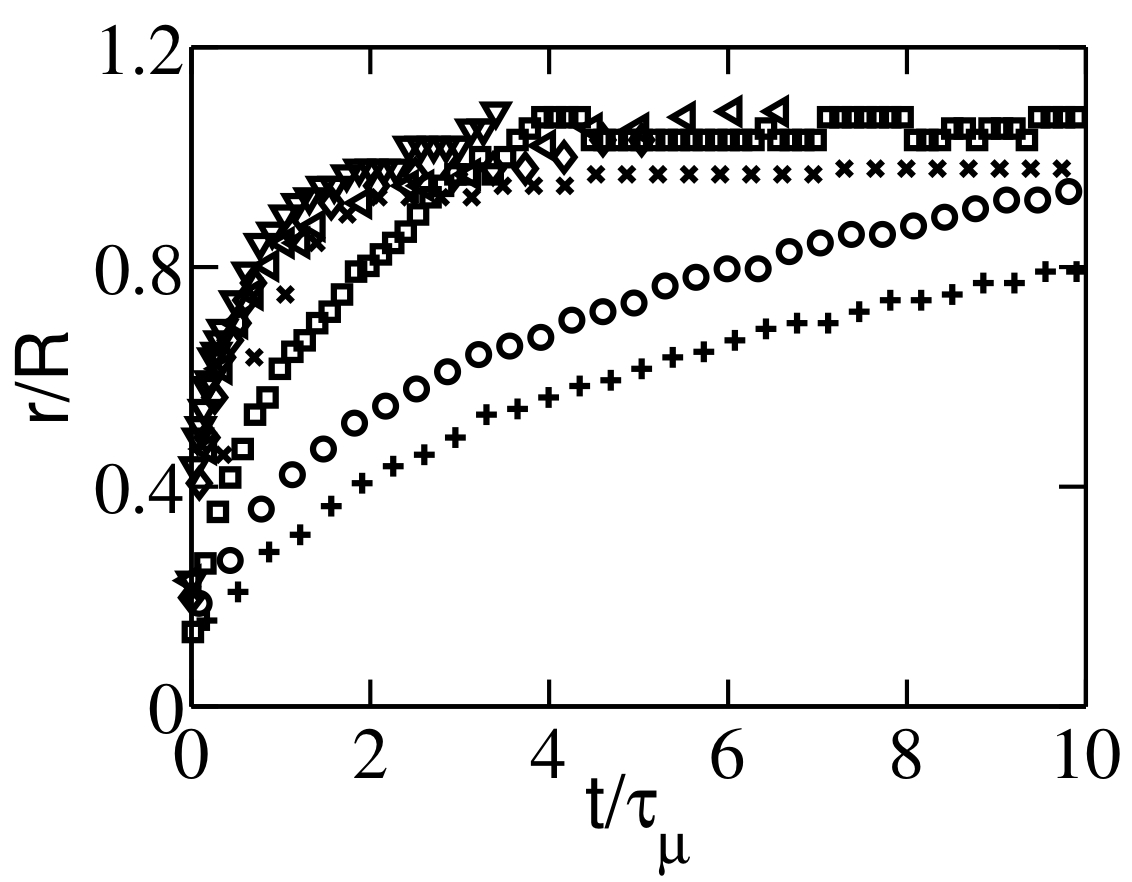}}
\label{fig:2b}
}
\caption{Apparent radius ($r$) scaled with a) an inertial time scale $\tau_{\rho}$ and b) a viscous time scale $\tau_{\mu}$. The data set has been reduced for clarity and the markers in Fig. 5 and Figs. S\ref{sub:fig1}, S\ref{fig:Appinertia}, S\ref{fig:Appviscous} denote the same data.}
\label{sub:fig1}
\end{figure}
An inertial ($\tau_{\rho}$) and a viscous ($\tau_{\mu}$) time scale are found to describe the drop deformation after contact with the oil film (Fig. 5), where Fig. S\ref{fig:Appdet} shows the time history of the drop shapes. A similar scaling law is also expected for the apparent radius ($r$). By scaling $r$ with the initial drop size ($R$) and time ($t$) with $\tau_{\rho}$ or $\tau_{\mu}$ two distinct spreading regimes appear (Fig. S\ref{sub:fig1}). The inertial scaling (Fig. S\ref{sub:fig1}a) shows that the data for $Oh<0.2$ follows nearly the same curve, where as $Oh>0.2$ the data remains scattered. If $r$ is instead scaled with $\tau_{\mu}$ the data for $Oh<0.2$ is reduced to nearly a single curve (Fig. S\ref{sub:fig1}b). The data in the transitional regime ($0.2 <Oh<2$) follows neither of the two scaling laws. 

\begin{figure}[ht!]
\centering
\subfigure[]{
{\includegraphics[width=0.45\linewidth]{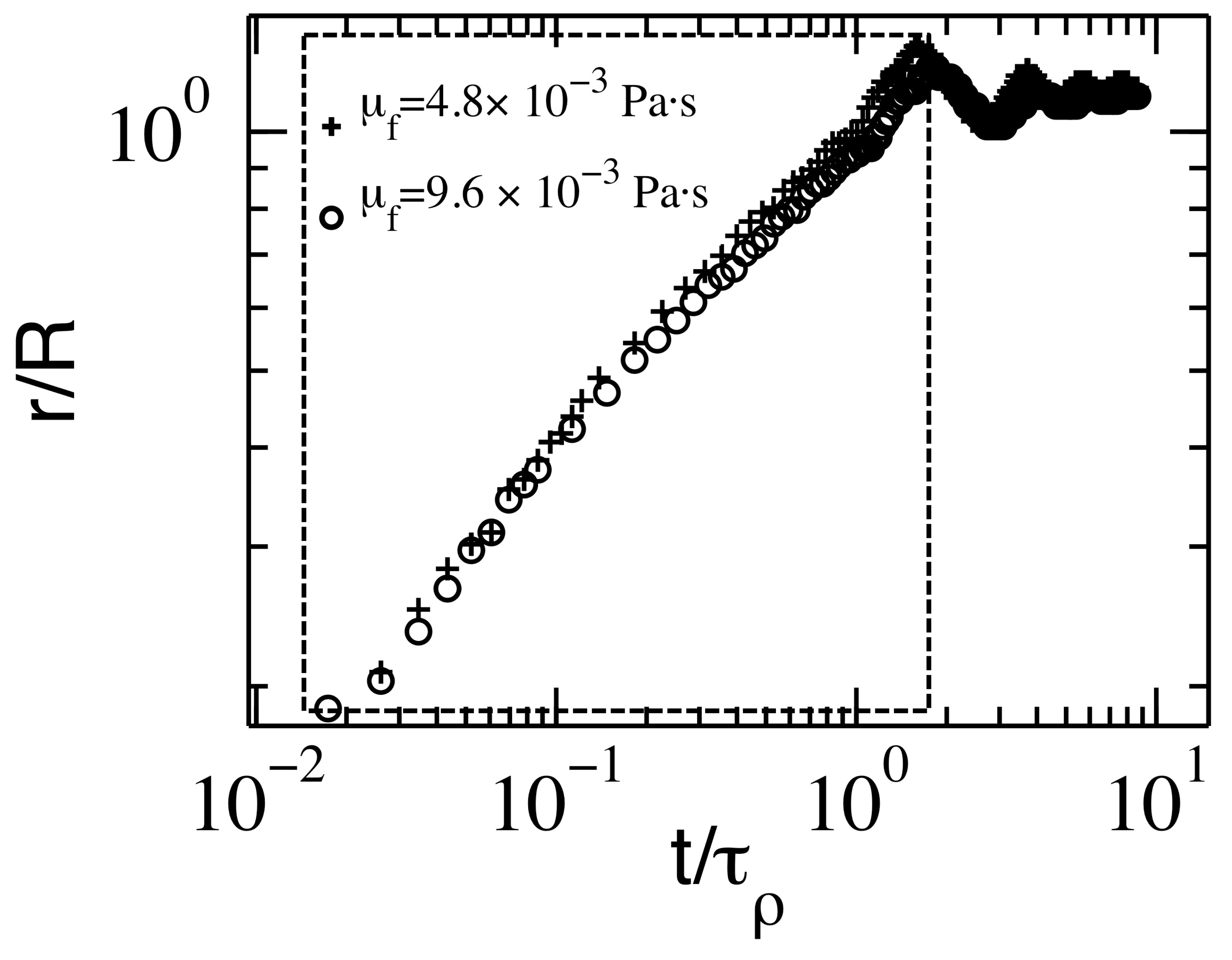}}
}
\centering
\subfigure[]{
{\includegraphics[width=0.45\linewidth]{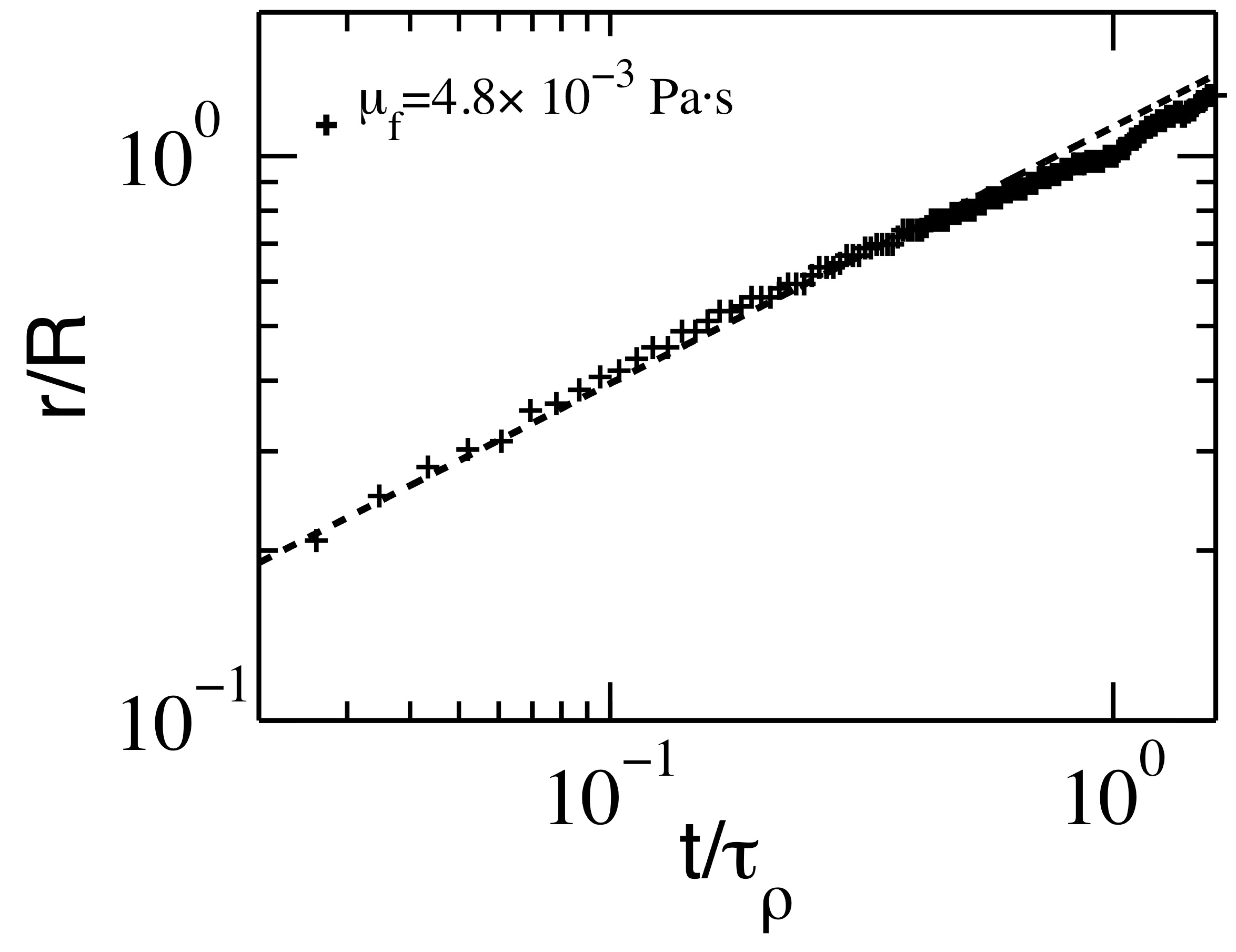}}
}
\centering
\subfigure[]{
{\includegraphics[width=0.45\linewidth]{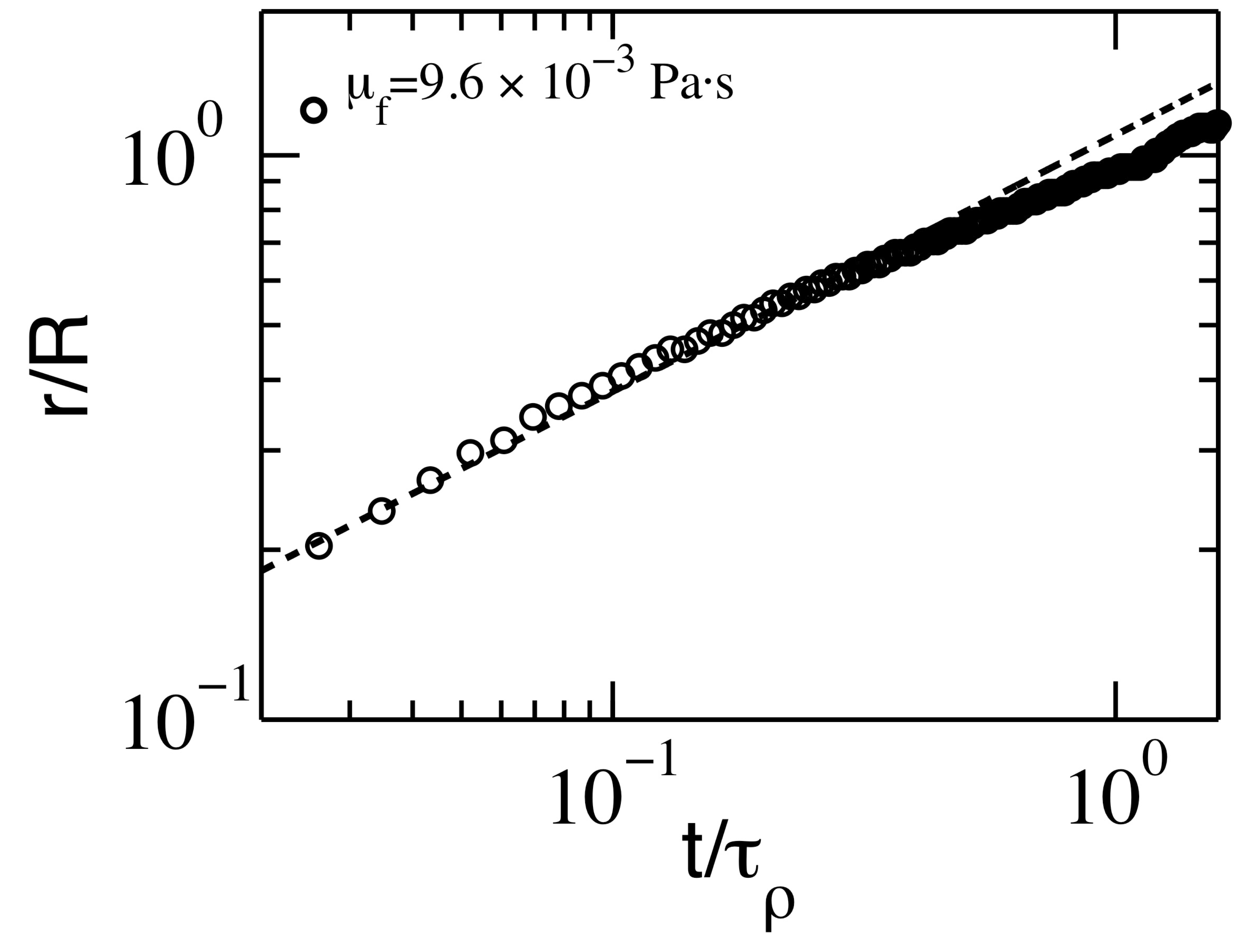}}
}
\caption{Determining the best fitted curve to describe the inertial scaling law for $r/R$. a) Non-dimensional log-log plot of the inertial scaling of the apparent radius. The data inside the dashed square is curve fitted onto a master curve to describe the power-law. b) Film viscosity $\mu_o=4.8\times 10^{-3} Pa~s$ and the best fit is represented by the dashed line, $r/R=1.13(t/\tau_{\rho})^{0.45}$. c) Film viscosity $\mu_o=9.6\times 10^{-3} Pa~s$ and the best fit is represented by the dashed line, $r/R=1.09(t/\tau_{\rho})^{0.45}$.}
\label{fig:Appinertia}
\end{figure}

\begin{figure}[ht!]
\centering
\subfigure[]{
{\includegraphics[width=0.48\linewidth]{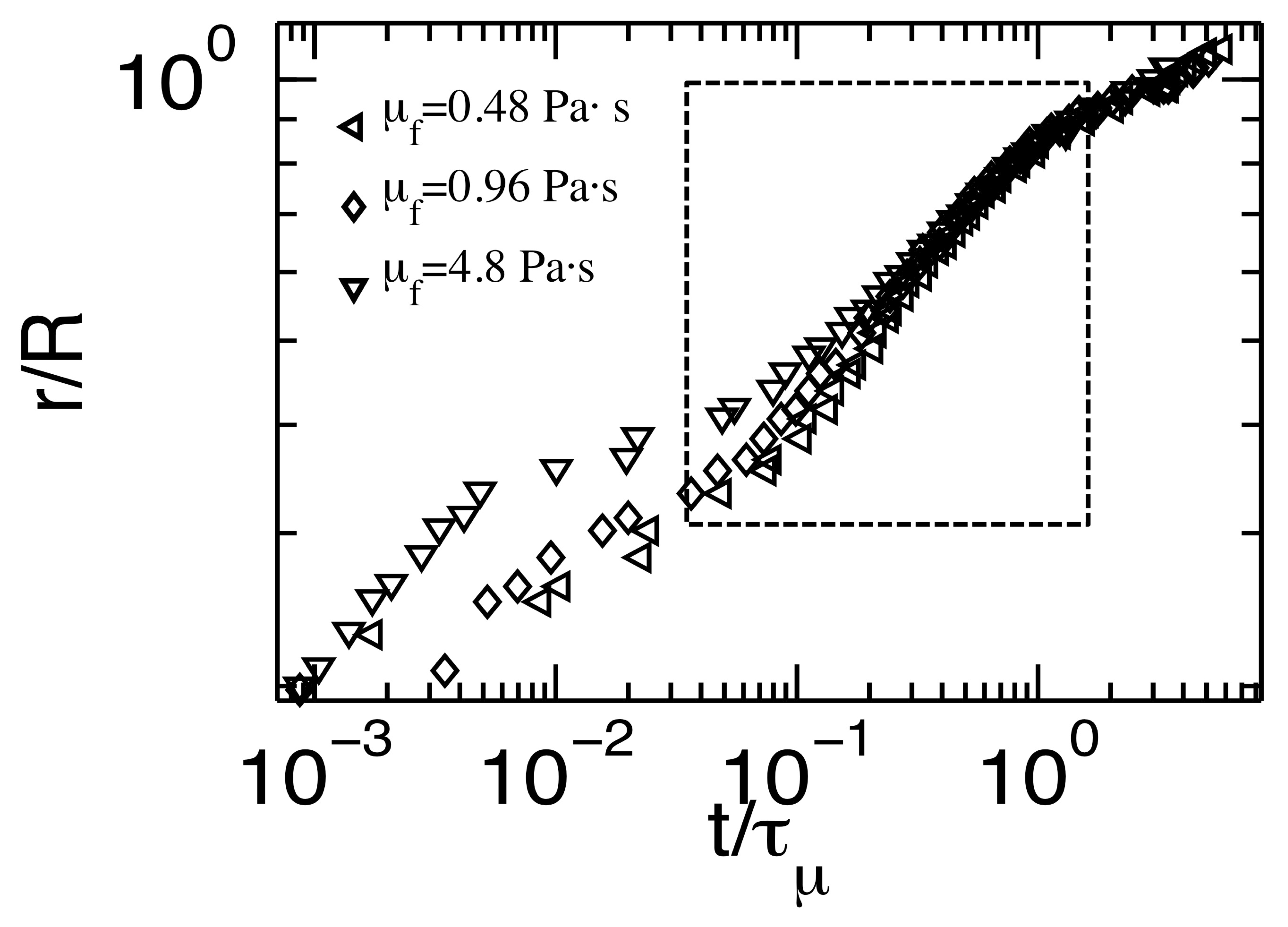}}
}
\centering
\subfigure[]{
{\includegraphics[width=0.45\linewidth]{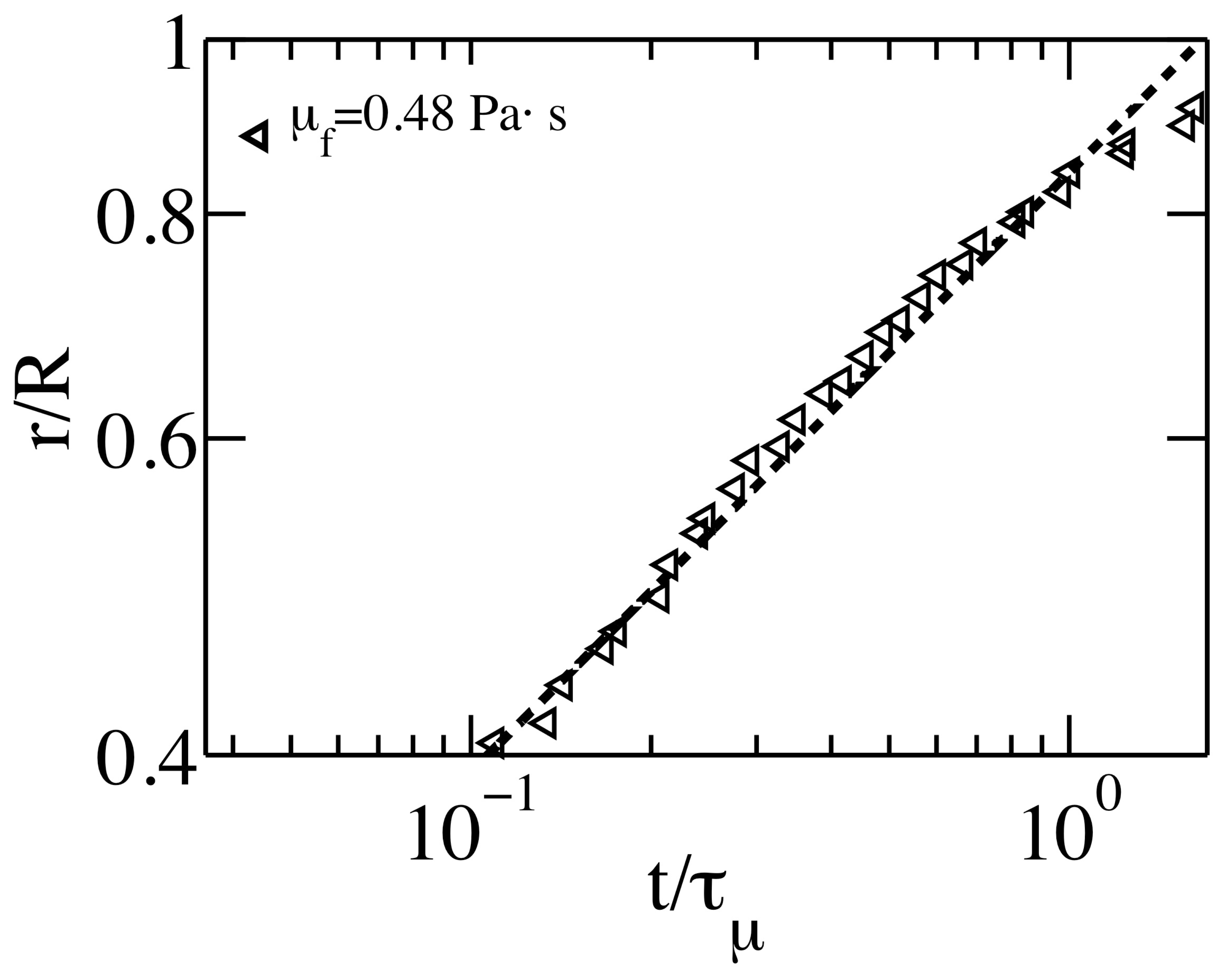}}
}
\centering
\subfigure[]{
{\includegraphics[width=0.45\linewidth]{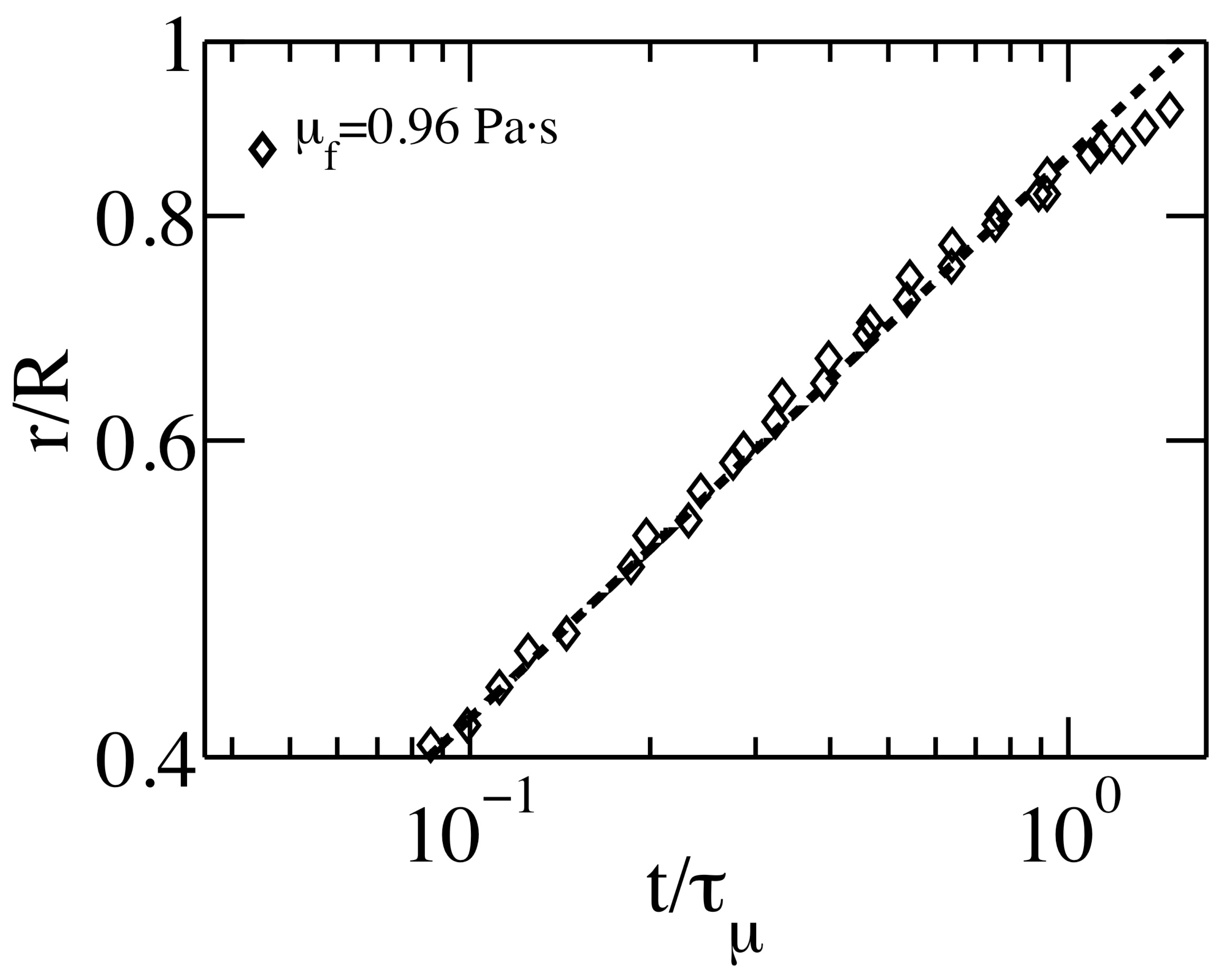}}
\label{fig:2b}
}
\centering
\subfigure[]{
{\includegraphics[width=0.45\linewidth]{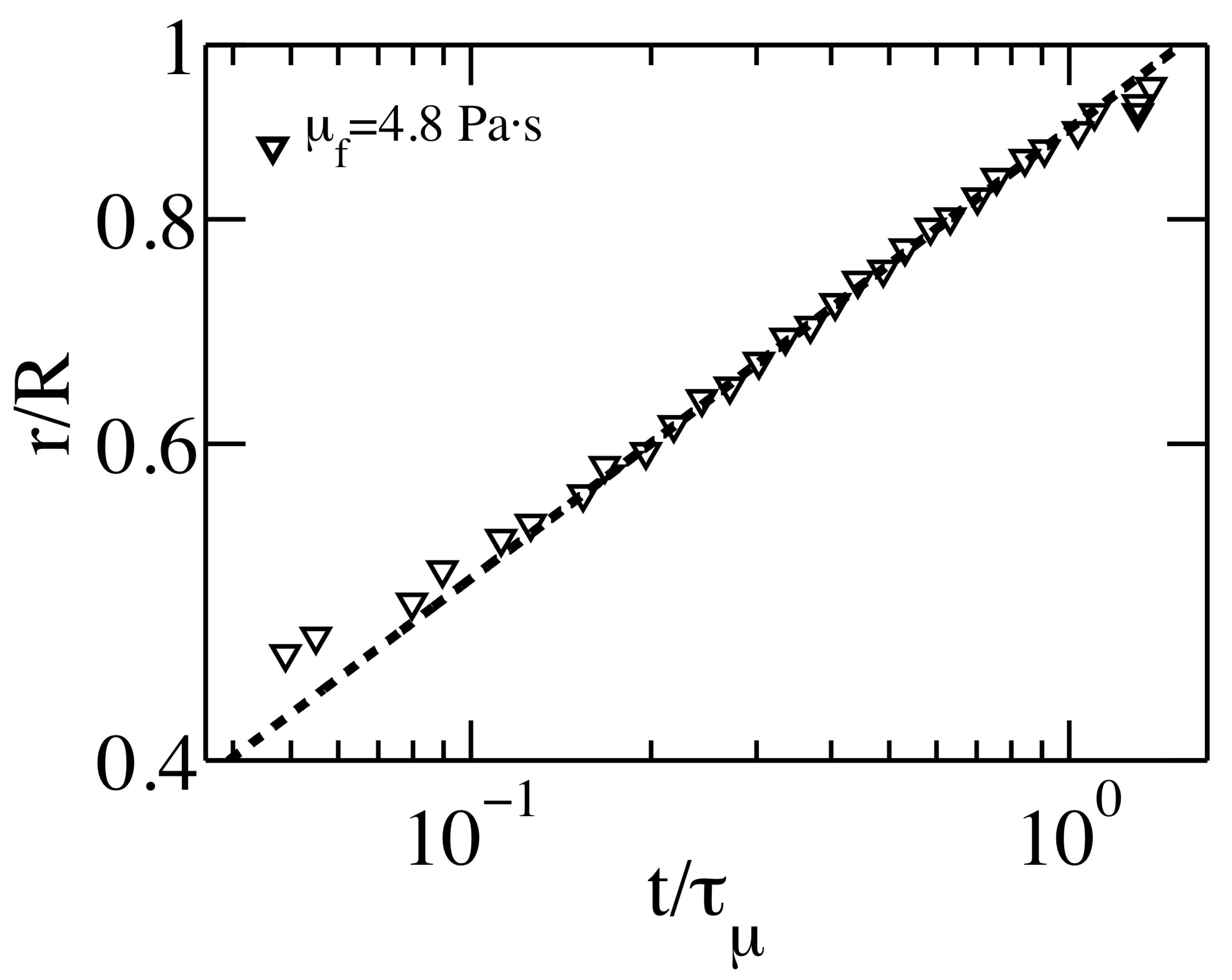}}
}
\caption{Determining the best fitted curve to describe the viscous scaling law for $r/R$. a) Non-dimensional log-log plot of the viscous scaling of the apparent radius. The data inside the dashed square is curve fitted onto a master curve to describe the power-law. b) Film viscosity $\mu_o=0.48~Pa~s$ and the best fit is represented by the dashed line, $r/R=0.84(t/\tau_{\mu})^{0.33}$. c) Film viscosity $\mu_o=0.96~Pa~s$ and the best fit is represented by the dashed line, $r/R=0.86(t/\tau_{\mu})^{0.31}$. d) Film viscosity $\mu_o=4.8~Pa~s$ and the best fit is represented by the dashed line, $r/R=0.90(t/\tau_{\mu})^{0.25}$.}
\label{fig:Appviscous}
\end{figure}

A log-log plot of the non-dimensional apparent radius ($r/R$) in Figs. S\ref{sub:fig1}a,b shows that it follows a power-law like behavior (Figs. S\ref{fig:Appinertia}, S\ref{fig:Appviscous}). We fit a master curve through each data set to determine the range of the exponents and the pre-factors for the two regimes. Since the experimental data show that these two regimes are bounded within a certain time span we only use the data in these regions for the fitting, which is illustrated by dashed square in Figs. S\ref{fig:Appinertia}a, S\ref{fig:Appviscous}a. Each data set inside the dashed area is fitted onto a curve of the form $r/R=a(t/\tau_{\rho})^{b}$ or $r/R=a(t/\tau_{\mu})^{b}$, where $a$ and $b$ are constants determined by the best fit. For clarity, the data in the transitional regime ($0.2<Oh<2$) has been omitted below. 

The inertial regime is defined before drop detachment from the needle ($t/\tau_{\rho}<2$). Each data set is fitted by a single curve (Fig. S\ref{fig:Appinertia}), which is represented by the dashed line. We find the the inertial regime to be represented by a curve $r/R\approx 1.1(\frac{t}{\tau_{\rho}})^{0.45}$, where the pre-factor varies between 1.09 -- 1.13 and the exponent is $\approx 0.45$.

A similar fitting procedure as described above is applied for the viscous regime (Fig. S\ref{fig:Appviscous}). The viscous scaling law shows that there might be different slopes for $r/R$ during the spreading of the thin film (Fig. S\ref{fig:Appviscous}a). We define the viscous regime within the time span $0.1<t/\tau_{\mu}<1$. We note, however, that a much smaller exponent for the curve is observed at early ($t/\tau_{\mu}<0.05$) and late ($t/\tau_{\mu}>1.0$) times. The dynamics at early and late times is also believed to be dominated by viscous forces. As the greatest measurement uncertainty is at $t/\tau_{\mu}<0.05$, we do not find it is meaningful to extract the pre-factor or the exponent of a fitted master curve in this region. We focus here on $r/R$ in the time span $0.1<t/\tau_{\mu}<1$, where $r/R$ follows a curve $r/R\approx0.87(\frac{t}{\tau_{\mu}} )^{0.3}$. The pre-factors and exponents for the fitted master curve varies between 0.84 - 0.90 and 0.25 - 0.33, respectively. 

\clearpage

\section{Measurement of re-wetting hole dynamics after lubricant rupture}
\begin{figure}[ht!]
\centering
\begin{overpic}[width=1.0\linewidth]{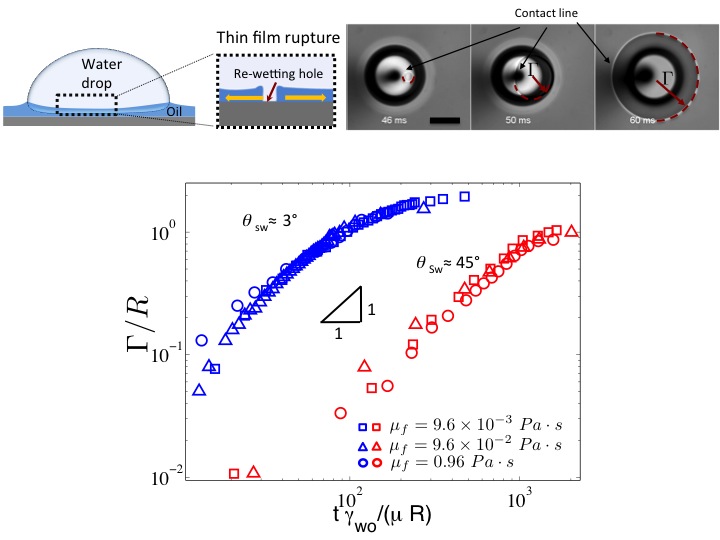}
\put(25,55){a)}
\put(75,55){b)}
\put(50,-2){c)}
\end{overpic}

\caption{a) Sketch of the initial formation of the water hole as the oil film punctures. b) Optical microscopy image of the growth of the wetting water hole in time. The scale bar is $200~\mu m$, $h_i=8~\mu m$, $\mu_o=9.6\times 10^{-3}~Pa\cdot s$ and $R=0.48~mm$. c) Growth of the re-wetting water hole ($\Gamma$) in time for different film viscosities and equilibrium contact angle.}
\label{fig:dewet}
\end{figure}

The thin oil film that separates the drop and the solid substrate becomes unstable for $\theta_{sw}<90^{\circ}$. Once the film punctures, a hole of water ($\Gamma$) comes in contact with the substrate that expands radially as water wets the substrate Fig. S\ref{fig:dewet}a. To visualize the re-wetting of water on the glass plate we visualize the interface dynamics through an optical microscope through the glass plate from below, which allow us to track the expanding hole. In Fig. S\ref{fig:dewet}b we illustrate how the expanding hole is extracted from the experiments. The place of the nucleation site varies for each individual experiment. The radius ($\Gamma$) of the wetting hole is extracted in time for different film viscosities ($\mu_o$) and equilibrium angles ($\theta_{sw}$). The speed of the expanding hole is in concordance with observations reported for de-wetting experiments by \textit{Redon et al.} [\textit{Phys. Rev. Lett.}, \textbf{11}, 016301 (1991)] Additional experiments were also performed in an oil bath, without any noticeable difference in the opening speed of the re-wetting water hole.

\section{Surface chemistry controlled water collection}
\begin{figure}[ht!]
\centering
\includegraphics[width=1.0\linewidth]{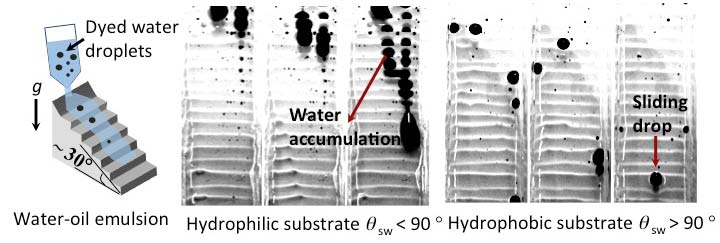}
\caption{a) Sketch of the experiment that demonstrates phase separation of an oil emulsion on a lubricated ($\mu_o=9.6\times 10^{-3} Pa\cdot s$) staircase. A water-oil emulsion is poured over a lubricated staircase which is b) hydrophilic and c) hydrophobic, where water is colored by a food dye to distinguish the two different phases. b) Collection of oil puddles at the hydrophilic staircase. c) Water drops are not pinned at the hydrophobic staircase.}
\label{fig:Appstairs}
\end{figure}

Another implication of knowing the lubricant stability is that it allow design that can promote separation of phases in oil emulsions. One such example is illustrated in Fig. S\ref{fig:Appstairs}, where a water-oil emulsion is poured over a lubricated staircase with different chemical treatment. Water is colored by food dye to better identify the two different phases. Water is collected at the hydrophilic staircase, but the emulsion remains well mixed on the hydrophobic staircase (Fig. S\ref{fig:Appstairs}c).

\end{document}